\documentclass{article}
\usepackage{graphicx}
\usepackage{amsmath}

\begin{document}

\title{On the Energy-Momentum Densities of the Cylindrically Symmetric Gravitational Waves}
\author{Ali Havare$^{1}$
\footnote{Corresponding author. E-mail: alihavare@mersin.edu.tr},
Mustafa Salt\i $^{1}$, Taylan Yetkin$^{2}$ \\
\linebreak $^{1}$Department of Physics, Mersin University, 33342,
Mersin-Turkey\\
$^{2}$Department of Physics, \c{C}ukurova University, 01330, Adana-Turkey}
\maketitle

\begin{abstract}
In this study, using M\o{}ller and Tolman prescriptions we calculate energy
and momentum densities for the general cylindrically symmetric spacetime
metric. We find that results are finite and well defined in these complexes.
We also give the results for some cylindrically symmetric spacetime models.
\end{abstract}

\section{Introduction}
In general relativity a large number of researches has been
devoted considerable attentions to the problem of finding the
conserved quantities, shuch as energy and momentum, associated
with various spacetimes. Following Einstein's original
pseudotensor for energy-momentum many expressions have introduced
in the literature\cite{1,2,3,4,5,6,7,8,9,10}.

Einstein's energy momentum complex, used for calculating the
energy in general relativistic system, was followed by many prescriptions,
e.g. M\o{}ller's, Papapetrou's, Landau-Liftshitz's (LL), Tolman's and
Weinberg's. Einstein's, Papapetrou's and LL's complexes give meaningful
results when the line-element is transformed to the cartesian coordinates.
However for the M\o{}ller's and Tolman's prescriptions it is not necessary
to use of Cartesian coordinates. In the literature Virbhadra and his collaborators
have considered many spacetimes and have shown that several energy-momentum
complexes give the same and acceptable results\cite{11}.
Aguirregabiria \textit{et al.}\cite{12} showed that several energy-momentum complexes
give the same result for any Kerr-Shild class metric. Recently, Chang \textit{et al.}\cite{13}
showed that every energy-momentum complex can be associated with a
particular Hamilton boundary term, therefore the energy-momentum complexes
may also be considered as quasi-local.

The paper organized as follow. In Section 1 We shortly introduce the generic, spatially
homogeneous rotating spacetimes and give some kinematical variables for this model.
In Section 3 and Section 4 we compute the energy and momentum densities for the line-element
which is introduced in Section 2 by using M\o{}ller's and Tolman's prescriptions, respectively.
Finally in Section 5 we give the results and conclusions. Throughout this paper we choose units
such that $G=1$ , $c=1$ and follow the convention that Latin indices take values from
$0$ to $3$ and Greek indices take values $1$ to $3$ otherwise stated.

\section{Spatially Homogeneous Rotating Spacetimes}

After the pioneering works of Gamow\cite{14} and G\"{o}del\cite{15}, the idea of global
rotation of the universe has became considerably important physical aspect
in the calculations of general relativity. A line-element which describes
spatially homogenous universes with rotation but with zero shear can be
given as follow
\begin{equation}
ds^{2}=-dt^{2}+dr^{2}+A(r)d\phi ^{2}+dz^{2}-2B(r)dtd\phi .  \label{1}
\end{equation}
In this study we discuss some special solutions of the M\o{}ller
and Tolman prescriptions for the energy and momentum densities by using the
general model given in (\ref{1}). For this model the matrices of the $g_{\mu
v}$ and $\Delta ^{2}g^{\mu v}$ are found as follow
\begin{equation}
\left(
\begin{array}{cccc}
-1 & 0 & -B & 0 \\
0 & 1 & 0 & 0 \\
-B & 0 & A & 0 \\
0 & 0 & 0 & 1
\end{array}
\right) ,\text{ \ \ \ \ }\left(
\begin{array}{cccc}
-A & 0 & -B & 0 \\
0 & \Delta ^{2} & 0 & 0 \\
-B & 0 & 1 & 0 \\
0 & 0 & 0 & \Delta ^{2}
\end{array}
\right)
\label{2}
\end{equation}
where $\Delta (r)=\sqrt{A+B^{2}}$. Using the Christoffel symbols which are
defined by
\begin{equation}
\Gamma _{\mu v}^{\alpha }=\tfrac{1}{2}g^{\alpha \beta }(g_{\mu \beta
,v}+g_{v\beta ,\mu }-g_{\mu v,\beta }). \label{3}
\end{equation}
we obtain the nonvanishing Christoffel symbols as
\begin{eqnarray}
2\Delta ^{2}\Gamma _{\mu v}^{0} &=&BB^{\prime }(\delta _{\mu }^{0}\delta
_{v}^{1}+\delta _{\mu }^{1}\delta _{v}^{0})+(AB^{\prime }-BA^{\prime
})(\delta _{\mu }^{1}\delta _{v}^{2}+\delta _{\mu }^{2}\delta _{v}^{1}) \\
2\Gamma _{\mu v}^{1} &=&B^{\prime }(\delta _{\mu }^{0}\delta _{v}^{2}+\delta
_{\mu }^{2}\delta _{v}^{0})-A^{\prime }\delta _{\mu }^{2}\delta _{v}^{2} \\
2\Delta ^{2}\Gamma _{\mu v}^{2} &=&(A^{\prime }+BB^{\prime })(\delta _{\mu
}^{1}\delta _{v}^{2}+\delta _{\mu }^{2}\delta _{v}^{1})-B^{\prime }(\delta
_{\mu }^{0}\delta _{v}^{1}+\delta _{\mu }^{1}\delta _{v}^{0}).
\label{4}
\end{eqnarray}
One can introduces the tetrad basis as follow
\begin{equation}
\theta ^{0}=dt+Bd\phi ,\text{ \ \ \ }\theta ^{1}=dr,\text{ \ \ \ \ }\theta
^{2}=\Delta d\phi ,\text{ \ \ \ \ }\theta ^{3}=dz.  \label{5}
\end{equation}
With the comoving tetrad formalism, the kinematical variables of this model
can be expressed solely in terms of the structure coefficients of the tetrad
basis which are defined as \cite{16}
\begin{equation}
d\theta ^{\alpha }=\tfrac{1}{2}c_{\beta \gamma }^{\alpha }\theta ^{\beta
}\Lambda \theta ^{\gamma }. \label{6}
\end{equation}
By taking the exterior derivatives of the tetrad basis which we
introduced and the following kinematics formulas \cite{17}
\begin{equation}
\begin{array}{ll}
\text{four-acceleration vector:} & a_{i}=c_{i0}^{0} \\
\text{vorticity tensor:} & \omega _{ij}=\frac{1}{2}c_{ij}^{0} \\
\text{expansion(deformation) tensor:} & \theta _{ij}=\frac{1}{2}%
(c_{i0j}+c_{j0i}) \\
\text{expansion scalar:} & \theta =c_{01}^{1}+c_{02}^{2}+c_{03}^{3} \\
\text{vorticity vector:} & \omega ^{1}=\frac{1}{2}c_{23}^{0},\text{ \ \ }%
\omega ^{2}=\frac{1}{2}c_{31}^{0},\text{ \ \ }\omega ^{3}=\frac{1}{2}%
c_{12}^{0} \\
\text{vorticity scalar:} & \omega =\frac{1}{4}[%
(c_{23}^{0})^{2}+(c_{31}^{0})^{2}+(c_{12}^{0})^{2}]^{1/2} \\
\text{shear tensor:} & \sigma _{ij}=\theta _{ij}-\frac{1}{3}\theta \delta
_{ij}.
\end{array} \label{7}
\end{equation}
we find the following quantities for Eq.(\ref{1})
\begin{equation}
\omega^{1} = \omega^{2} = 0,\,\,\,\ \omega ^{3}=\frac{B^{\prime }}{2\Delta }, \,\,\,\ a_{i}=\theta
_{ij}=\sigma _{ij}=0 \label{8}
\end{equation}
where prime indicates derivative with respect to $r$. The model given in Eq.(\ref{1})
is shear-free and has non-vanishing vorticity. We also note that this
frame has a vanishing four-acceleration. The line-element (\ref{1}) can be
reduced to well-known spacetime models in the literature by imposing some
conditions on metric functions:
\begin{enumerate}
\item  \textit{The Rebou\c{c}as spacetime}\cite{19} \newline
Defining $A(r)=-(1+3c^{2}),$ $B(r)=2c$ with $c=\cosh 2r$  the line-element
\ref{1} reduces to
\begin{equation}
ds^{2}=-dt^{2}+dr^{2}-(1+3c^{2})d\phi ^{2}+dz^{2}+4cdtd\phi \label {9}
\end{equation}

\item  \textit{The Som-Raychaudhuri spacetime}\cite{19} \newline
Defining $A(r)=r^{2}(1-r^{2}),$ $B(r)=r^{2}$ the line-element
\ref{1} reduces to
line-element
\begin{equation}
ds^{2}=-dt^{2}+dr^{2}+r^{2}(1-r^{2})d\phi ^{2}+dz^{2}-2r^{2}dtd\phi \label{10}
\end{equation}

\item  \textit{The Hoenselaers-Vishveshwara spacetime}\cite{19}\newline
Defining $A(r)=-\frac{1}{2}(c-1)(c-3),$ $B(r)=c-1$ with $c=\cosh r$
the line-element \ref{1} reduces to
\begin{equation}
ds^{2}=-dt^{2}+dr^{2}-\frac{1}{2}(c-1)(c-3)d\phi ^{2}+dz^{2}-2(c-1)dtd\phi \label{11}
\end{equation}

\item  \textit{The G\"{o}del-Friedman spacetime}\cite{19}\newline
Defining $A(r)=s^{2}(1-s^{2}),$ $B(r)=\sqrt{2}s^{2}$ with $x^{\mu }=a%
\widetilde{x}^{\mu }$ the line-element \ref{1} reduces to
\begin{equation}
ds^{2}=-dt^{2}+dr^{2}+dz^{2}+s^{2}(1-s^{2})d\phi ^{2}-2\sqrt{2}s^{2}dtd\phi \label{12}
\end{equation}

where $x^{\mu }=(t,r,\theta ,z)$, $\widetilde{x}^{\mu }=(\eta ,\varepsilon
,\phi ,y)$, $a=(1,1)\times 10^{25}m$ and $s=\sinh r$

\item  \textit{The Stationary G\"{o}del spacetime}\cite{15}

Defining $A(r)=-e^{2ar}/2, $ $B(r)=e^{ar}$, $a=$ constant,
the line-element \ref{1} reduces to
\begin{equation}
ds^{2}=-(dt+e^{ar}d\theta )^{2}+dr^{2}+\frac{1}{2}(e^{ar}d\theta )^{2}+dz^{2} \label {13}
\end{equation}
\end{enumerate}

\section{Energy and Momentum Densities in M\o{}ller's Prescription}
Since the M\o{}ller complex is not restricted to the use of
Cartesian coordinates we can perform the computations in $t,$ $r,$
$\theta ,$ $z$ coordinates. The computations in these coordinates
are easier compared to those in $t,$ $x,$ $y,$ $z$ coordinates.

The following is the M\o{}ller's energy-momentum complex $\Theta
_{a}^{b}$ \cite{21}:
\begin{equation}
\Theta _{i}^{k}=\frac{1}{8\pi }\chi _{i,l}^{kl}  \label{14}
\end{equation}
where
\begin{equation}
\chi _{i}^{kl}=-\chi _{i}^{lk}=\sqrt{-g}[g_{in,m}-g_{im,n}]g^{km}g^{nl} \label{15}
\end{equation}
here $\Theta _{0}^{0}$ is the energy density, $\Theta _{\alpha
}^{0}$ are the momentum density components. The M\o{}ller
energy-momentum obeys the local conservations laws
\begin{equation}
\partial _{k}\Theta _{i}^{k}=0 \label{16}
\end{equation}
The energy-momentum components are expressed by
\begin{equation}
P_{i}^{M}=\iiint \Theta _{i}^{k}dx^{1}dx^{2}dx^{3} \label{17}
\end{equation}
where $\mu _{\beta }$ are the components of a normal vector over an
infinitesimal surface element $dS$. $P_{\alpha }^{M}$ give the momentum
components $P_{1}^{M},$ $P_{2}^{M},$ $P_{3}^{M}$ and $P_{0}^{M}$ gives the
energy. For the line element (\ref{1}) the non-vanishing components  of $\chi _{i}^{kl}$ are
found as
\begin{eqnarray}
\chi _{0}^{01} &=&-\frac{BB^{\prime }}{\Delta } \label {18} \\
\chi _{2}^{01} &=&\frac{BA^{\prime }-AB^{\prime }}{\Delta }  \label {19} \\
\chi _{1}^{01} &=&\chi _{3}^{01}=0 \label{20}
\end{eqnarray}
Using these results in equation (\ref{3}), we obtain energy and momentum
densities in M$\phi $ller's prescription as
\begin{eqnarray}
\Theta _{i}^{0} &=&\frac{1}{8\pi}\chi _{i,1}^{01} \\ \label{21}
\Theta _{j}^{0} &=& 0. \label{22}
\end{eqnarray}
where $i = 0, 2$ and $j = 1,3$. By using the metric functions
which are given in Section 2 it is easy to calculate the energy and momentum densities.
The results obtained in M\o{}ller's prescription are given in Table \ref{tab:table1}.

\section{Energy and Momentum in Tolman's Prescription}
Tolman's energy and momentum complex \cite{5} is given by
\begin{equation}
\Phi_{i}^{k}=\frac{1}{8\pi }\Lambda_{i,l}^{kl}  \label{23}
\end{equation}
where $\Phi _{0}^{0}$ and $\Phi _{\alpha }^{0}$ are the energy and momentum
components. We have
\begin{equation}
\Lambda_{i}^{kl}=\sqrt{-g}[-g^{pk}\Omega _{ip}^{l}+\tfrac{1}{2}g_{k}^{i}g^{pm}%
\Omega _{pm}^{i}] \label{24}
\end{equation}
with
\begin{equation}
\Omega _{jk}^{i}=-\Gamma _{jk}^{i}+\frac{1}{2}g_{j}^{i}\Gamma _{mk}^{m}+%
\frac{1}{2}g_{k}^{i}\Gamma _{mj}^{m}. \label{25}
\end{equation}
Also, the energy and momentum complex $\Phi _{i}^{k}$ satisfies
the local conservation laws
\begin{equation}
\partial _{k}\Phi _{i}^{k}=0. \label{26}
\end{equation}
The energy and momentum in Tolman's prescription are given by
\begin{equation}
P_{i}=\iiint \Phi _{i}^{k}dx^{1}dx^{2}dx^{3} \label{27}
\end{equation}
where $n_{\alpha }$ are the components of a normal vector over an
infinitesimal surface element $dS$. $P_{\alpha }$ give the momentum
components $P_{1},$ $P_{2},$ $P_{3}$ and $P_{0}$ gives the energy. For the
line element (\ref{1}) the non-vanishing components of $U_{i}^{kl}$ are found as
\begin{eqnarray}
\Lambda_{0}^{01} &=&\frac{A^{\prime }+BB^{\prime }}{2\Delta } \label{28} \\
\Lambda_{2}^{01} &=&\frac{BA^{\prime }-AB^{\prime }}{2\Delta } \label{29} \\
\Lambda_{1}^{01} &=&\Lambda_{3}^{01}=0 \label{30}
\end{eqnarray}
using these results in Eq.(\ref{25}), we obtain energy and momentum
densities in Tolman's prescription
\begin{eqnarray}
\Phi_{i}^{0} &=&\frac{1}{8\pi}\Lambda_{i,1}^{01} \label{31} \\
\Phi_{j}^{0} &=& 0. \label{32}
\end{eqnarray}
where $i = 0, 2$ and $j = 1, 3$. Again with the metric functions' definitions
which are given in Section 2 it is easy to calculate the energy and momentum densities.
The results obtained in Tolman's prescription are given in Table \ref{tab:table1}.

\begin{table}
\begin{tabular}{|l||l|l||l|l|}
\hline
&\multicolumn{2}{l|}{M\o{}ller's Prescription}
 &\multicolumn{2}{l|}{Tolman's Prescription}\\
\cline{1-5}
Space-time/Components &$\Theta_{0}^{0}$& $\Theta_{2}^{0}$ & $\Phi_{0}^{0}$&$\Phi_{2}^{0}$ \\
\hline\hline
Rebou\c{c}as             & $-\frac{s}{\pi}$  & $-\frac{6sc^{2}}{\pi}$
                         & $-\frac{s}{2\pi}$ & $-\frac{3sc^{2}}{\pi}$ \\
Som-Raychaudhri          & $-\frac{r}{2\pi}$ & $-\frac{r^{3}s}{2\pi}$
                         & $-\frac{r}{4\pi}$ & $-\frac{r^{3}s}{4\pi}$ \\
Hoenselaers-Vishveshwara & $-\frac{s}{4\sqrt{2}\pi}$ & $\frac{s(c-1)}{4\sqrt{2}\pi}$
                         & 0 & $\frac{s(c-1)}{8\sqrt{2}\pi}$\\
G\"{o}del-Friedman       & $-\frac{cs}{\pi}$ & $-\frac{\sqrt{2}cs^{3}}{\pi}$
                         & 0 & $-\frac{cs^{3}}{\sqrt{2}\pi}$\\
Stationary G\"{o}del     & $-\frac{a^{2}e^{ar}}{8\pi}$   & $-\frac{a^{2}e^{2ar}}{4\sqrt{2}\pi}$
                         & 0 & $-\frac{a^{2}e^{2ar}}{8\sqrt{2}\pi}$ \\
\hline
\end{tabular}
\caption{Energy and momentum densities for the specific space-time
models. The $c$ and $s$ values in the table are given in Sec. 2.}
\label{tab:table1}
\end{table}
\section{Discussions}
We used prescriptions of M\o{}ller and Tolman to calculate energy
and momentum densities for a general form of cylindrically symmetric spacetime.
We also gave the energy and momentum densities for the special cases of genaral
cylindrically symmetric spacetime. We found that M\o{}ller's prescription give
twice the Tolman's prescription,
\begin{equation}
P_{i}^{M}=2P_{i}^{T}. \label{33}
\end{equation}
Also, for the Hoensalaers-Vishveshwara, G\"{o}del-Friedman and G\"{o}del spacetimes
we found that energy components is vanishing in the Tolman's prescription. For this spacetimes
one has to perform calculations in cartesian coordinates.

\end{document}